\documentclass[a4paper,11pt]{article}
\usepackage{epsfig}
\usepackage{graphicx}
\usepackage{natbib}
\usepackage{a4wide}
\usepackage{rotating}
\usepackage{longtable}

\begin{document}
\newcommand{\logg}{$\log g$}
\newcommand{\teff}{$T_{\rm eff}$}
\newcommand{\loggp}{$\log g(1)$}
\newcommand{\loggs}{$\log g(2)$}
\newcommand{\teffp}{$T_{\rm eff}(1)$}
\newcommand{\teffs}{$T_{\rm eff}(2)$}
\newcommand{\feh}{[Fe/H]}
\newcommand{\av}{$A_{V}$}
\newcommand{\Msun}{$M_{\odot}$}

\noindent{\Large {\bf A study of supervised classification of Hipparcos variable stars using PCA and Support Vector Machines}}\\


\noindent{\large \bf{
P.G. Willemsen$^1$, L. Eyer$^2$}
}
 \vspace{0.2cm}

\noindent
(1) Sternwarte der Universit\"at Bonn, Auf dem H\"ugel 71, 53121 Bonn, Germany\\
(2) Observatoire de Gen\`eve, 51 ch. des Maillettes, 1290 Sauverny, Switzerland\\ 
\vspace{1mm}

Report no.: VSWG-PW-001
\vspace{1mm}

2005.06.15

\vspace{0.5cm}
\noindent {\bf Abstract}: We report on the automated classification of Hipparcos variable stars by a supervised classification algorithm known as Support Vector Machines. The dataset comprised about 3200 stars, each characterized by 51 features. These are the $B-V$ and $V-I$ colours, the skewness of the lightcurve, the median subtracted 10-percentiles and forty bins from the Fourier envelope of the lightcurve. We also tested whether the classification performance can be improved by using the most significant principal components calculated from this dataset.
We show that the overall classification performance (as measured by the fraction of true positives) on the original dataset is of the order of 62~\%. 
For about 9 of the 18 different variability classes, the classification accuracy is significantly larger than 60~\% (up to 98~\%).
Introducing principal components does not significantly improve this result. We further find that many of the different variability classes are not very distinct and possibly poorly defined, i.e. there exists a considerable class overlap. It is concluded that this `contamination' of the template set implies minimum errors and thus degrades the overall performance.

\tableofcontents

\section{Introduction}
\label{intro}

Given that many stars which Gaia will observe are variable and given that there are many different types of variability, a precise and efficient classification of these objects is very important. Recently, \citealt{Eyer04} tested an unsupervised Bayesian classifier to automatically extract the variability classes for about 1800 stars with observed lightcurves. 
Similarly, \cite{Evans04} used an (unsupervised) algorithm known as Self-Organizing Map (SOM) to classify variable stars. 

The classification performance naturally depends on the data set as well as on the algorithm used for the analysis. To get an overview of the principal capabilities of these different algorithms, we have tested the classification performance of a supervised method known as Support Vector Machine on Hipparcos variable stars. The data in this work is different from the above mentioned studies. However, it is ultimately foreseen to use a homogeneous set of variable stars and to test different algorithms on these data, similarly to the Blind Testing procedure as performed in the classification working group (ICAP, see e.g. \citealt{ICAPAB03}).

The classification performance and computational efficiency of supervised and unsupervised classification algorithms can sometimes be improved by reducing the dimensionality of the original dataset.
Especially for the case of time-dependent photometric data as being obtained from variable stars, the number of attributes which characterize such stars is rather large (see Section \ref{data} and also \citealt{Eyer04}).

A common tool to perform such a \textit{feature extraction} (the term `feature' here means the attributes or observations for a given star)
is the principal component analysis (PCA). Here, the dimensionality is reduced while retaining as much as possible the variation of the original data, i.e. one tries to isolate the most descriptive and discriminatory features in the data set. These new features are then used for the subsequent analysis. 

In PCA, one basically solves the eigenvalue problem of the covariance matrix (or correlation matrix, see Sect.\,\ref{ana}) of the data features. The eigenvector with the $k$th eigenvalue is the $k$th principal component (PC) and has the $k$th largest variance among all PCs. Note that the principal components are linear transformations of the original features. Moreover, all PCs are uncorrelated and the $k$th PC can be interpreted as the direction that maximizes the variation of the projections of the data points such that it is orthogonal to the $k -1$ other PCs.   

To see, whether we can help in the classification process, we did a detailed differential study on the classification performance based on the original data and PCA-based data. 

\section{Data set and data preprocessing}
\label{data}

The data set was basically compiled from the Hipparcos variable stars.
In total there are 11236 stars and 55 data entries (features) for each star.  
These are the Hipparcos number (HIP), the $B-V$  and $V-I$ colours, the number of measurements, the skewness or asymmetry ($asy$) of the lightcurve, the median subtracted 10-percentiles ($d1$ to $d9$), forty bins from the Fourier envelope ($f1$ to $f40$) and the object type (in case of a missing class, this is a blank).

\subsection{Object types}

In total, there are 4486 stars with known variability classes (types). However, several of these are subclasses (for example EA/DM) and another fraction is of uncertain nature (e.g. RRC?).

In order to make the data more homogeneous which is important for the subsequent classification work, all subclasses were assigned to their corresponding (main) classes (e.g. EA/DM $\rightarrow$ EA). Systems of uncertain nature and systems with combined classes (e.g. BY+UV) were rejected. Moreover, we rejected all stars with class assignments which belong to `supersets' of other classes. For example, class E is a superset of classes EA and EB, i.e. classes E and EA are not mutually exclusive.

Next, only those classes were chosen from the revised data set which had at least 40 stars as representatives. This absolute number of stars per class is somewhat arbitrary here and was mainly motivated by the requirements that there are several different object classes in the data set to be analysed where each of these must be represented many times in order to allow for a good classification.
    
    \begin{center}
\begin{table}[h!]
\centering
\begin{tabular}{c|c|c}
\cline{1-3}
\multicolumn{2}{c|}{Variability class}  & \multicolumn{1}{c}{} \\
\cline{1-3}
rejected & retained & \# of stars per class\\
\cline{1-3}
   BE    &   ACV & 170\\
   BY    &   ACYG  & 48\\  
   BY+UV &   BCEP  & 59 \\ 
   CEP   &  DCEP  & 188 \\
    CW   &   DSCT  & 111 \\
    CWA  & EA  &  472\\
    CWB  &  EB  &  324\\
    DCE  &  ELL &  47 \\
    DCEPS & EW   & 113\\
    E+ZAN & GCAS  & 198\\
    EA+BC &  I    & 517\\
   EA+DSC & L   &  356\\
   ELL+XF & M   &  190\\
   FKCOM  & RRAB  & 75\\ 
   NC     &   RS  & 68\\
   NL    &  SPB  & 91\\
   NL+ZZ &   SRA  & 42 \\
   NR    & SRB  & 148\\
 PVTEL   & \\
 RCB     &  \\
 RV     & &   \\
 RVA    & &  \\
 RVB     & &\\
 S       & & \\
 SARV    & & \\
 SDOR    & & \\
 SPB     & & \\
 SR+ZA   & & \\
   SR:/PN& & \\
   SRA+E & & \\
   SRC   & & \\
   SRD   & & \\
   SXARI & & \\
   SXPHE & & \\
  UV     & & \\
  WR     & & \\
  XNG    & & \\
  ZAND   & & \\
\cline{1-3}
\end{tabular}
\caption{The first column shows the variability classes which were rejected during the data selection process. The second and third columns show those classes which passed the selection criteria and the corresponding number of stars per class. In total there are 3211 stars in the selected sample.\label{sets}} 
\end{table}
\end{center}

\subsection{Colours}    
    
As mentioned above, the $B-V$ and $V-I$ colours were provided in the data set. Given that these colours are somewhat redundant, we tested if one of these should be omitted during the analysis, or if they should be combined to a single colour $B-I$ (and thus reducing the dimensionality). 
For this, we examined colour-colour diagrams such as shown in the lower plot of Fig.\,\ref{col}. It can be seen that the objects cannot be clearly separated from $B-V$ alone. For example, there are several objects of different variability classes at high $B-V$. The colour $V-I$ on the other hand seems to be a better indicator for the variability class, i.e. using only this colour in the analysis would probably be sufficient. However, there are several reasons why we would not want to rely on $V-I$ or $B-I$ only.

First, there is no real benefit in reducing the dimension of the dataset by one. The computational efficiency of a classification algorithm will not be significantly different in case that the dimension of the dataset was reduced to 51, instead of 52. At such high dimensions, these small changes can be ignored.

Second, most automated algorithms (at least SVMs and neural networks) can handle a possible redundancy in the dataset rather well by `internally' weighting these features during training via some free parameters.
At this point it should be noted that the MBP instrument on board Gaia will contain 14 different filters, sampling the wavelength region from $\sim$ 2000 to 12000 \AA. It is clear that there is much redundancy in such data, much more than the possible redundancy of the (broad-band) $B-V$ and $V-I$ colours in this work.  Recently, a PCA of these MBP data was performed and the new dataset (with 6 instead of 14 dimensions) was used in combination with a neural network for testing the parametrization performance. It was found that the parametrization performance did not change significantly (Willemsen 2005, unpublished) as compared to the original dataset with 14 dimensions. This is also in agreement with what was found in \cite{Bailer98}. 

Another reason why we would want to use both colours ($B-V$ and $V-I$) in the analysis is that we cannot exclude that the combination of these yield more information than a single colour. Indeed, a close inspection of the upper plot in Fig.\,\ref{col} shows that dwarf and giant stars can only be separated in the twodimensional colour space. 

Finally, we did limited classification tests with both colours and with $V-I$ alone. Since we did not find significant differences in the classification performances, we chose to use $B-V$ in addition to $V-I$ in the following.

\subsection{Cleaned features}
\label{feat}

\begin{figure}
\centering
\includegraphics[scale=0.6,angle=0]{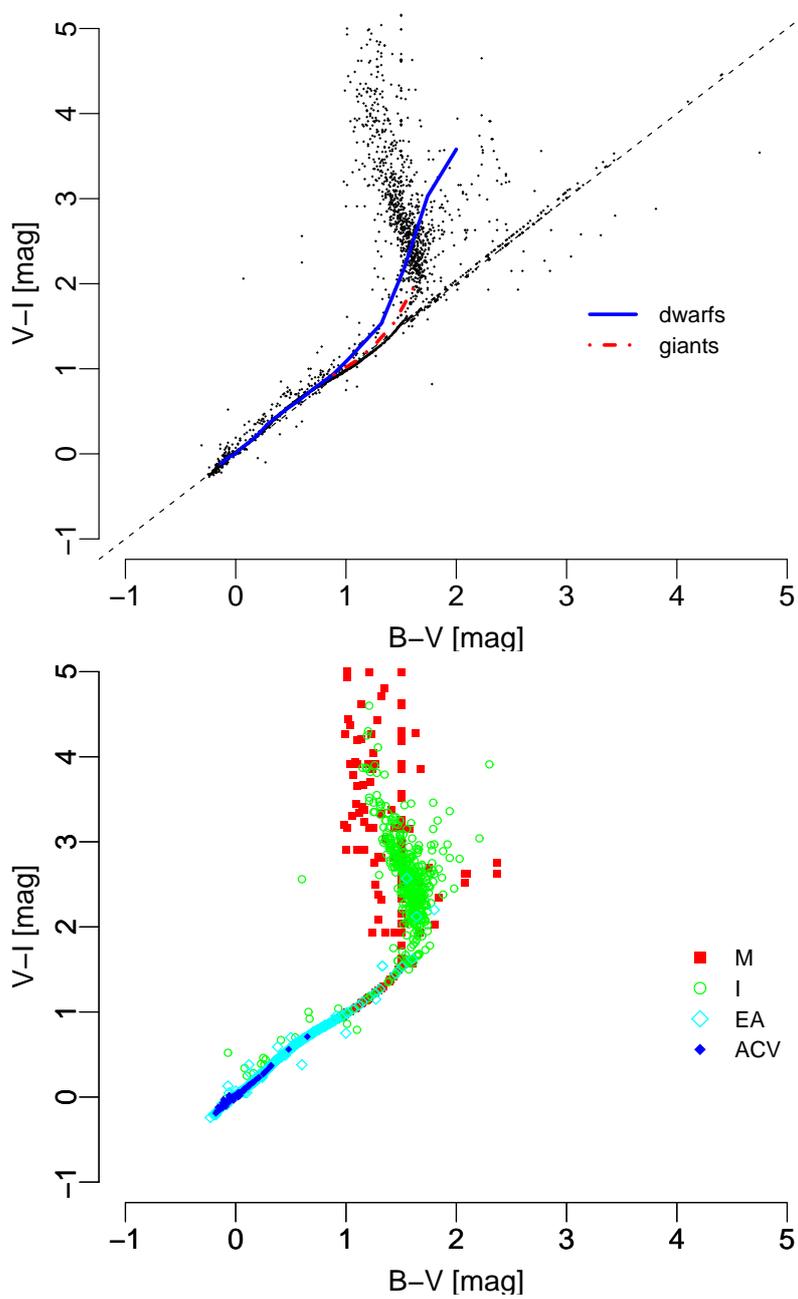}
\caption{The upper plot shows the colour-colour diagram for all stars with known variability classes. Shown as blue and red lines are the empirical colour-colour lines (\citealt{Allen20}) for dwarf and giant stars (for $B-V$ $\leq$ 1.6 mag), respectively. The black dashed line is the identity. It can be seen that the variable stars roughly follow the empirical trend of stable stars in that the difference between $B-V$ and $V-I$ becomes larger for redder objects, although there is a large scatter.  The lower plot shows the `cleaned' set (see Sect.\,\ref{feat}). As an example, the stars in four different variability classes are highlighted.
\label{col}}
\end{figure}

To allow for a comparison of what one would expect for normal (stable) stars, we overplotted the empirical colour-colour relations for dwarf and giant stars in Fig.\,\ref{col} (data taken from \citealt{Allen20}). It can be seen that there is a large scatter of the $V-I$ colour for very red objects and the general trend does not really follow the empirical curve for normal type stars. A possible explanation for this scatter (when compared to the colour-colour relation of normal stars) might be that the colours of certain variable stars must be weighted by the corresponding periods. For example, RR Lyrae type stars vary by several tenths of magnitudes in $B-V$ during one period, i.e. the mean colour can only be determined by taking the lightcurve into account. Moreover, the atmospheres of very evolved stars (e.g. Mira type stars or some irregularly pulsating variable stars) have different stellar opacities as compared to hydrogen burning stars. 

From the Figure we also note that there are very red objects which lie close to the identity line in the $B-V$ - $V-I$ plane. These are carbon stars (types C, R and N), which are mostly irregular or semi-regular variables. Since carbon stars will more easily be identifiable by the MBP photometry or the RVS spectra than by their lightcurves, we excluded these objects from the present analysis. 
We therefore rejected all objects with $B-V$ $>$ 2.5 and close to the identity line.
In a next step, we removed all obvious outliers in the colour-colour diagram by a) selecting only those objects with $V-I$ $<$ 5 mag and b) deleting all single points which deviate significantly from the major object distribution. 

The resulting colour-colour magnitude diagram after preprocessing the data file (class selection and colour inspection) is shown in the lower plot of Fig.\,\ref{col}.

\begin{figure}
\centering
\includegraphics[scale=0.7,angle=0]{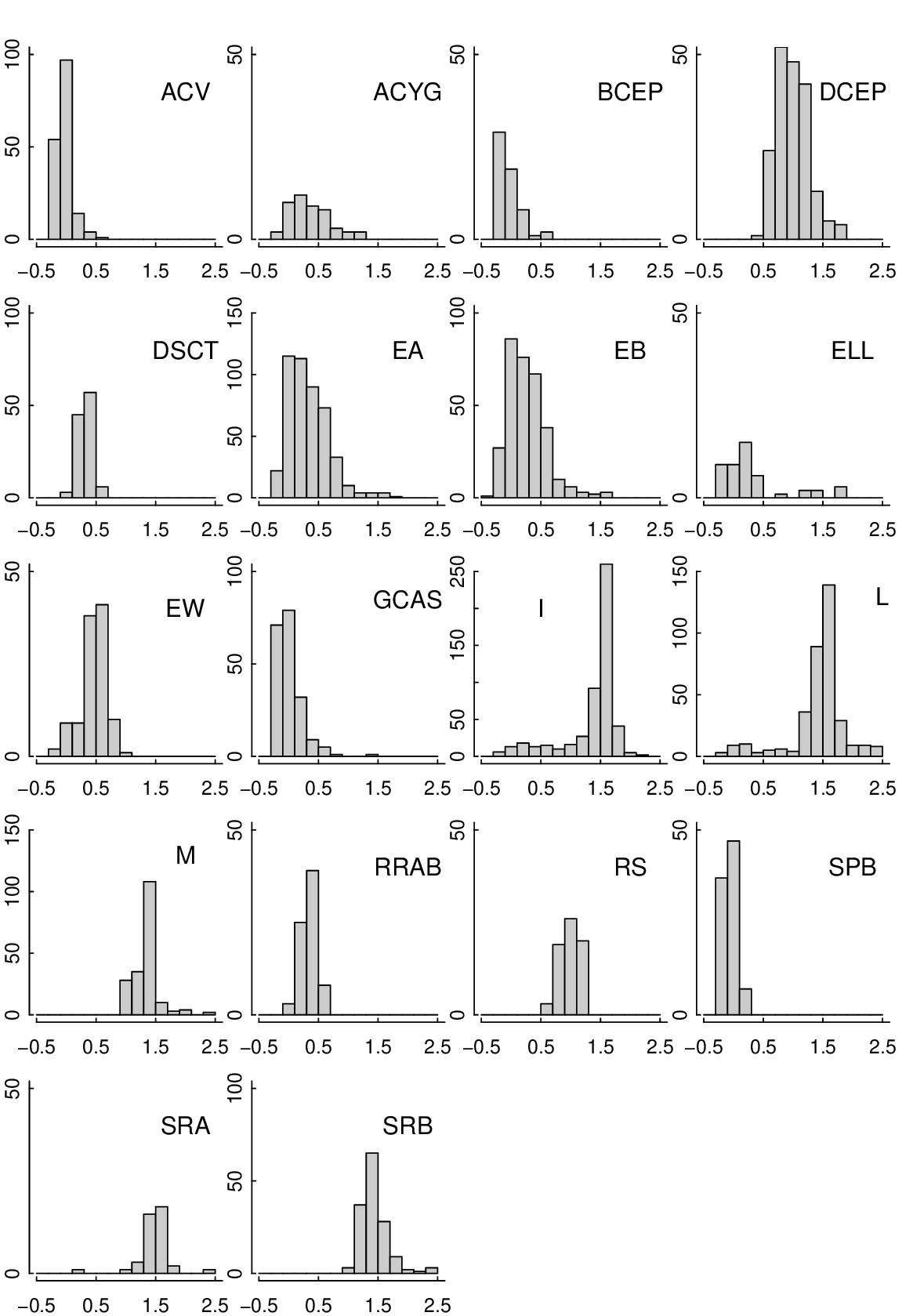}
\caption{ The $B-V$ colour distributions for the stars in the 18 different variability classes.  Note that the y axis has different scales for different classes, depending on the number of objects in each class.   
\label{BV}}
\end{figure}

A summary of the different variability classes and the corresponding numbers of objects in each class is given in Table\,\ref{sets}. In total there are 18 different variability classes which were chosen for the subsequent analysis. It should be noted that the number of objects per class differs by a factor up to $\sim$ 17. Depending on the classification algorithm (and the possible choice of class-weighting), this non-uniform distribution of objects might deteriorate the overall classification results. 

In Fig.\,\ref{BV} the $B-V$ colour distributions are shown for the stars in the different variability classes. It can be seen that the mean colours differ significantly for certain classes. It should be noted however, that the colour of a variable star need not be necessarily a good feature for representing the variability class.  
For example, the colours of eclipsing binaries are determined by the age and masses of the individual components, i.e. they essentially follow a 
\textit{random} distribution sampled by the components' masses and lifetimes. 
As a result, the intra-class spread of the colours can be larger for certain object classes. 
Likewise, RR Lyrae stars can have a colour spread of the order of $\Delta(B-V) \sim$1 mag, which could possibly have a deteriorating effect on the classification performance for such objects (the object is not well defined by a single colour measurement).

Summarizing, there are 3211 objects in 18 different variability classes in the final (`cleaned') data set.
For the PCA, 51 different features were used. These are the $B-V$ and $V-I$ colours, the asymmetry parameter ($asy$), the median subtracted percentiles $d1$ to $d9$ (without $d5$ since this is per definition always zero) and the 40 Fourier envelope components ($f1$ to $f40$).

\section{PCA Analysis} 
\label{ana}

The results of the PCA will depend on the scaling of the inputs. This is of special relevance if the inputs are of largely different nature, since e.g. larger values will tend to have larger variances. In a first step, all inputs were thus scaled by subtracting the mean from each value, where the mean is the average across each feature column.

Moreover, it is sometimes useful and necessary to define principal components via the \textit{correlation} matrix instead of the covariance matrix. This is simply the covariance of a pair of features scaled by the corresponding product of the standard deviations. Note that this procedure is equivalent to computing the principal components from the original features after being scaled to have unit variance. 

Using the correlation matrix instead of the covariance matrix is usually preferred in case that there are large differences between the variances of the features. If this is not corrected for (via scaling the variances), the first principal component will be dominated by the largest features with the largest variances. 
For our data sample, the minimum variance is found to be 0.01 (for $d4$) and the maximum variance is found to be 1.65 (for $V-I$), i.e. the variances of the different features differ by roughly two orders of magnitude. Scaling the variances should thus be appropriate for this kind of problem. 
For comparison, we also present the results for unscaled features (note that the mean was always subtracted).

\begin{figure}
\centering
\includegraphics[width=8cm,height=16cm,angle=-90]{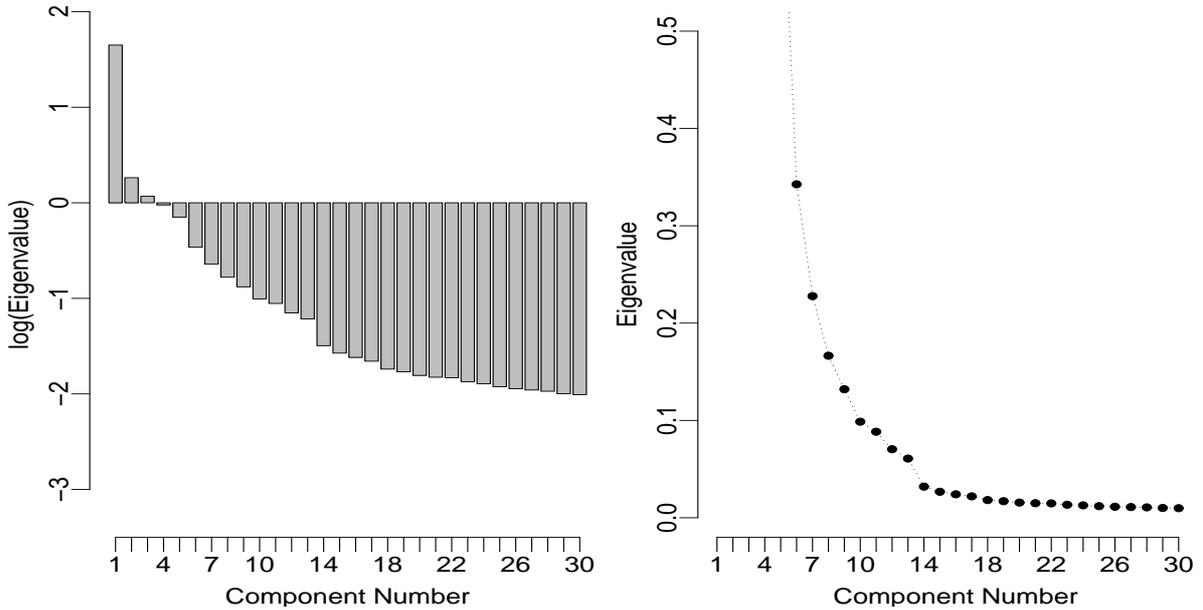}
\caption{Results for the PCA \textbf{with} variance scaling. Shown in the left plot is a histogram of the \textit{logarithmic} eigenvalues for the first 30 (out of 51) principal components. The right plot shows the eigenvalues but rescaled in order to emphasize the transition region from the important (larger) eigenvalues to the smaller eigenvalues (representing random variations) which tend to lie along a straight line. 
\label{pca11a}}
\end{figure}
\begin{figure}%
\centering
\includegraphics[width=8cm,height=16cm,angle=-90]{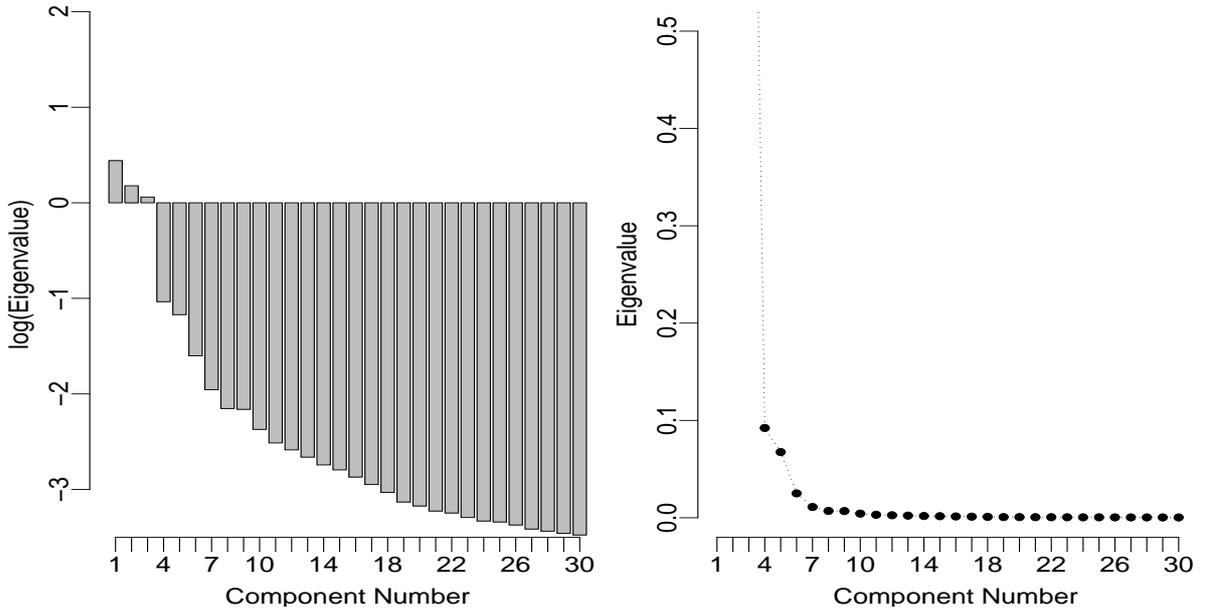}
\caption{The same as in Fig.\,\ref{pca11a} but \textbf{without} variance scaling. 
\label{pca12a}}
\end{figure}

Figs.\,\ref{pca11a} and \ref{pca12a} show the eigenvalues of the principal components as so called `screeplots' for the PCA with and without variance scaling, respectively. 

From the Figures we note that the eigenvalue of the first component is much larger in value if the variances are scaled ($\sim$ 45) than in case without variance scaling ($\sim$ 3).


Moreover, the distribution of the eigenvalues falls off more steeply in case that there is no variance scaling. In Fig.\,\ref{pca11a} the contributions of random variations (represented by small eigenvalues) starts at the $\sim$ 14th component, while for the case that the variance is not scaled (Fig.\,\ref{pca12a}) the curve seems to level off at the $\sim$ 7th component.

\begin{figure}
\includegraphics[scale=0.56,angle=-90]{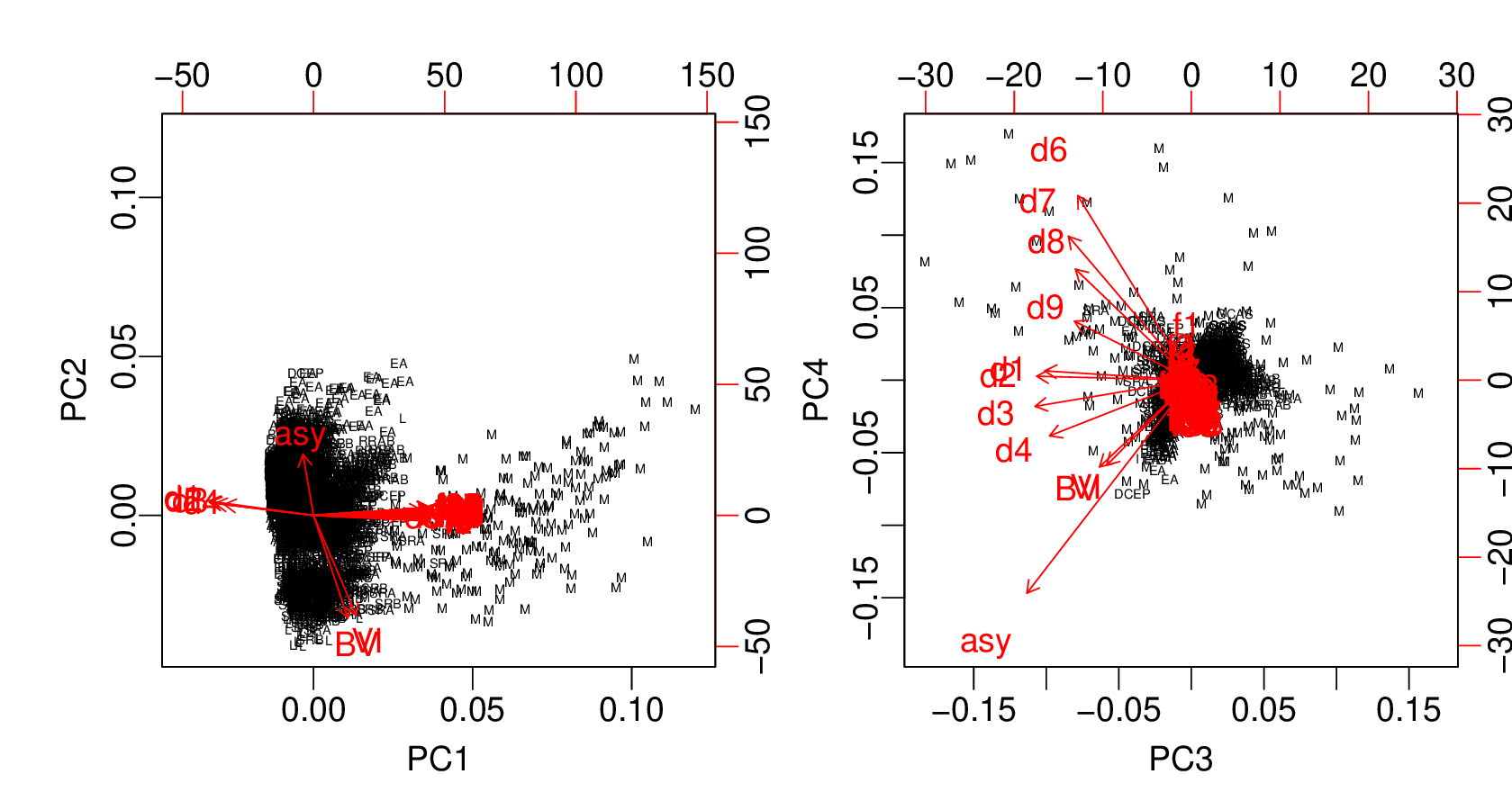}
\caption{ Biplots of the results presented in Fig.\,\ref{pca11a}, i.e. for the PCA \textbf{with} variance scaling. The left plot shows the plane of the first (x-axis) and second (y-axis) principal component, while the right plot shows the plane of the third and fourth PC. The lower and left scale refer to the (scaled) observations, the upper and right scale refer to the arrows which represent the original variables from which the PCs were computed. Given the strong correlation of the Fourier bins, the arrows for these features are all overlapping. The same is true for certain percentile features.
\label{pca11b}}
\end{figure}%
\begin{figure}
\includegraphics[scale=0.56,angle=-90]{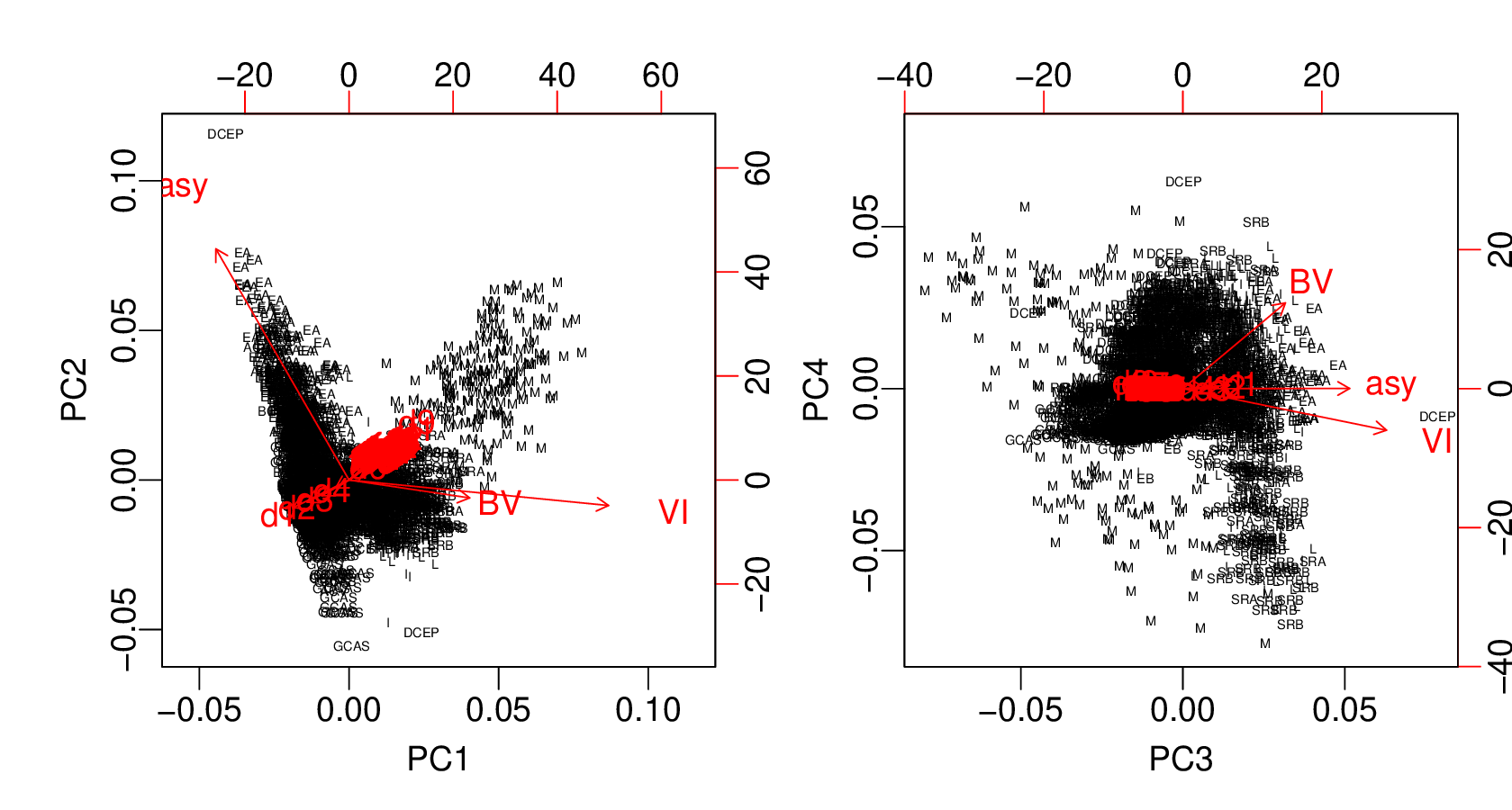}
\caption{Biplots for the data in Fig.\,\ref{pca12a}, i.e for the PCA \textbf{without} variance scaling. See Fig.\,\ref{pca11b} for explanations.  
\label{pca12b}}
\end{figure}%

The same results are also shown in Fig.\,\ref{pca11b} (\textbf{with} variance scaling) and Fig.\,\ref{pca12b} (\textbf{without} variance scaling) but now as so called `biplots'. A biplot combines information about the principal components with that of the observations. The observations are represented by the object types (e.g. `M' for Mira) in the plane formed by the chosen principal components. In the Figures these are the first and second PC (left panel) and the third and fourth PC (right panel). Note that only  a plot made up of the first two PCs is a biplot in the strict (traditional) sense. The coordinates of the observations are shown in the lower and left margins, while the coordinates on the upper and right axes refer to the arrows or vectors. The arrows represent the original variables from which the PCs were computed. Note that the scale of the arrows is different from that of the original observations and that the scales for the left and right plots differ, too.

One can read several important facts from such a biplot. At first, the orientation of the arrows with respect to the PC plane gives an indication of the importance of a particular feature to the PC.  For example, for the PCA \textbf{with} variance scaling (Fig.\,\ref{pca11b}) we see that the $B-V$ and $V-I$ features are almost parallel to the axis of the second PC, i.e. these features contribute much to the second component. Likewise, the 10-percentiles ($d1$ to $d9$) and the Fourier envelope bins contribute mostly to the first component. 
The angles between the vectors show their correlation. For example, the relatively small angle between the Fourier envelope bins in Fig.\,\ref{pca11b} shows a strong correlation between these features. Similarly, $asy$ and $B-V$ show a negative correlation in this plot. The observation that the arrows of $B-V$ and the Fourier bins are almost perpendicular is due to the fact that these features only weakly correlate (as expected). 

For a more detailed introduction to biplots, see e.g. \cite{Rossiter05} and especially \cite{Venables02}.

Comparing the biplots for the two cases (with/without variance scaling), we note that the first component (PC1) is mostly made up of $B-V$ and $V-I$ in case \textbf{without} variance scaling (Fig.\,\ref{pca12b}), while in case of variance scaling (Fig.\,\ref{pca11b}), PC1 mostly constitutes of $d1$ to $d9$ and the Fourier bins $f1$ to $f40$. 
Similarly, the third component resulting from the PCA with variance scaling (right plot in Fig.\,\ref{pca11b}) is mostly made up by $d1$ to $d4$. From the relative lengths of the arrows in this plot, it can be seen that the Fourier bins (with short arrows) do not contribute much, neither to the third nor to the fourth PC. These features mostly contribute to PCs which are perpendicular to the plane formed by the third and fourth PC (i.e. PC1). 

Clearly, the larger absolute values of $B-V$, $V-I$ and $asy$ result in larger variances and these are reflected in the contributions to the principal components. This demonstrates that the scaling of the variances is appropriate and necessary for this analysis.

From Figs.\,\ref{pca11b} and \ref{pca12b} we can further see that certain object classes scatter more in the principal component planes than others. Especially the Mira, DCEP and EA classes include objects with large variations. Generally, it appears that the Mira type stars can be better separated in that they form a single cluster (of objects) in the PC1-PC2 plane in Fig.\,\ref{pca11b}.


\subsection{PCA Stopping criteria}

Our goal is to compute new variables which represent a maximum of important and a minimum of unimportant information. 
In general, principal components with large eigenvalues represent a large proportion of the total variance while those with small eigenvalues represent random variations.

The difficulty is now to decide, which components are important and which are not. Unfortunately, there does not exist a common (theoretical) criteria to decide how many of the principal components should be included in the subsequent analysis. Indeed, most of the `rules' are informal and ad-hoc. For a very good overview of the different methods see e.g. \cite{Jackson93}.

The inspection of the eigenvalues for increasing component numbers (e.g. Fig.\,\ref{pca11a}) is a very straightforward and simple method of determining the optimal number of principal components. Since the small eigenvalues tend to lie on a straight line, any deviation from this line (towards larger values) will give us an indication which components are of importance and which are not. This, however, is complicated in case that there are no obvious breaks or if there are multiple breaks. 


Another very simple possibility to choose the optimal number of PCs is to include all components up to some previously (and arbitrarily) defined proportion of the total variance. For example, we could choose all components which comprise 98~\% of the total variance. According to \cite{Jackson93}, this approach is favoured by some statisticians but at the same time strongly rejected by others who find it unfounded and unreliable.

Better criteria use Bootstrap procedures or a `parallel' analysis (\citealt{Franklin95}) which involves a set of random data from which one can deduce the significance of the eigenvalues determined from the original data set. 
Since these approaches are computationally rather expensive, one should probably favour one of the simpler criteria for this stage of the testing procedure.

For the present work, we chose that number of components for which the eigenvalues deviate `significantly' from the straight line.
Based on Fig.\,\ref{pca11a} (with variance scaling), this is the subset made up of the first 14 components. The classification results based on this new set will be compared to that based on the original 51 input features. 

At this point it should be mentioned that the traditional choice of the first few principal components (representing most of the variation in the data) need not be the best one. Indeed, as was shown in \cite{Yeung01} for clustering gene expression data, there can exist sets of PCs which can yield much better performances and which are not made up of the first components. 
Concerning the clustering algorithm used for classification, it is possible that it yields a better performance, if at least some amount of noise is present in the data set. For several optimization problems, noise can act as an (additional) regularization tool, preventing the algorithm of getting stuck in local minima (see e.g. \citealt{ICAP04}).
Whether the results found in \cite{Yeung01} also apply to our classification problem must be checked for different classification and clustering algorithms and for different combinations of principal components. 


\section{Training and validation set}

Since Support Vector Machines are supervised classification algorithms, we need to build a training set in addition to a validation set.
For this, we randomly shuffled the complete set and in a next step devided it disproportionally into a training sample with 2140 templates and a validation sample with 1071 variable stars. 
The distribution of the different variability classes in each of these samples naturally follows the overall distribution of the stars in the complete set, i.e. there are different numbers of objects per variability class. This should be kept in mind when analyzing the results.


\section{Classification by Support Vector Machines}

For the classification we used an algorithm known as Support Vector Machine (SVM). In the following we will only briefly introduce the major concepts of SVMs since the underlying theory is beyond the scope of this report. The interested reader is referred to \cite{INTROSVM} and especially \cite{LEARNSOFT}. SVMs have a sound theoretical basis developed by Vapnik and Chervonenkis (see e.g. \citealt{Vapnik95} and \citealt{Vapnik98}) in the framework of the Statistical Learning Theory and Structural Risk Minimization.

As many other algorithms, SVMs optimize some parameters during learning (in terms of neural networks called weights)  according to some measure of performance or training error.
The major difference between a neural network (NN) and SVM or other learning algorithms is this measure of performance (also called cost function or risk) which results in different optimization strategies.  In this sense, SVMs can be simply seen as an alternative learning procedure when e.g. compared to the optimization strategies of NNs. 

Generally, a SVM separates classes by finding an optimal hyperplane which divides the training data by a maximal margin. In cases where the inputs cannot be linearly separated, the SVM performs an implicit mapping of the training data into a highdimensional feature space where a linear separation should be possible. Note that a linear separating hyperplane in a highdimensional feature space results in  a nonlinear separating hypersurface in the original input space. The solution of this approach is found in a quadratic optimization problem with inequality constraints which can be solved by a Lagrange function with well defined properties. Note that the convex error surface guarantees that the training procedure finds the global minimum in a finite number of steps (using a sensible gradient-descent algorithm). This is different from e.g., NN's error surfaces (here multi-layer perceptrons) which can have many local minima. `Support vectors' are those training data inputs which lie closest to the optimal separating hyperplane and which therefore define the decision boundary.

The application of SVMs requires tuning (adjusting) of  controlling parameters which define the approximation and generalization performance. We evaluated the optimum values by ten-fold cross validation on the training set.

For the present simulations we used the $R$-default $e1071$ package.

\section{Classification Results}

In this Section we present the classification results based on the
different datasets, i.e. for the original data set and for data based
on the first 14 principal components.

The results are summarized in terms of so called confusion tables,
which show the true objects in a given class versus the predicted
class.

The confusion tables permit to estimate the false positives and false
negatives for the classification of every variable type. They allow to get a grasp of the classification reliability and also to
understand where are the possible confusions between different specific types.

It should be noted though that the numbers of objects in the
validation set (and naturally in the training set, too) are rather
small for certain classes. This can make a direct comparison of the
classification performances for the different classes rather
difficult. Moreover, we face the problem of low-number statistics,
i.e. the results are probably not significant in certain cases and for
certain classes.
 
Table \ref{origohne} shows the results for the original data for SVMs,
while Tables\,\ref{pca14ohne} is for SVMs trained on data based on the
first 14 principal components.

A quick comparison of the classification performances for the
different datasets can be done by summing up the diagonal elements
(the true positives) in the confusion matrix and dividing the sum by
the total number of objects in the validation set (1071, see above).
These overall performances are given below each table.

\subsection{Original versus principal-component-based dataset}

A comparison of Tables\,\ref{origohne} and \ref{pca14ohne} shows that there is no significant improvement in using principal components. Indeed, we find that the overall classification performance (measured by the fraction of true positives) slightly decreases if principal components are used (from $\sim$ 62\% to 60 \%). 
Introducing PCA in order to decrease the number of dimensions in the training set does not necessarily yield better results, i.e. the gains of the lossy compression do not exceed the losses.

\subsection{Results for specific classes}

In the following, we will concentrate on Table\,\ref{origohne}, i.e. results obtained from SVMs trained on the original dataset with 51 features. We will not attempt to comment on each individual class/misclassification but rather want to highlight the major results and trends. 

Overall, we note that the classification performance strongly varies for the different classes.  For example, for class M we find a correct classification rate (true positives) of $\sim$ 98~\%, while for class ELL we find 0~\%. 

For class ACV (Alpha Canum Venaticorum) variables we find a correct classification rate of $\sim$ 79 \%. From the confusion table we see that the majority of false classifications (5 out of 53) in this group can be found in class GCAS (eruptive irregular variables). Likewise, the majority of the misclassifications in class GCAS are of type ACV (15 out of 69 objects in class GCVS are classified as ACV). 

The correct classification rates for object types ACYG and BCEP are rather low ($\sim$ 57 and 23 \%, respectively). It should be noted though, that the absolute number of objects per class is small (7 and 13) so that these results are possibly not representative in statistical terms. For the astrophysically important class of DCEP objects (Delta Cep variables) however, we find a high rate of true positives (89 \%) similar to the objects of class DSCT (Delta Scuti) with a correct classification rate of $\sim$ 90 \%.

For classes EA and EB (Algol type EA, Beta Lyrae-type  EB) we find that a rather high rate of class EB stars is misclassified as EA stars (about 55 \%). Similarly, 11 out of 28 objects of class EW (W Ursae Majoris-type) are classified as class EB stars. All three classes belong to the eclipsing binary systems and it is therefore possible that they share several similarities in their characteristics which makes it difficult to distuinguish between these objects.

Classes I (irregular variables) and L (slow irregular variables) both show a high rate of misclassifications. Per definition, class I belongs to the superset of eruptive variables, while class L belongs to the pulsating stars. A closer look at the definition of class I objects reveals that these are irregular variables with unknown features of light variations and spectral types. According to the GCVS (\citealt{GCVS}), class I is a `very inhomogeneous group of objects'. Similarly, objects of class L are slow irregular variables and these stars are `often attributed to this type because of being insuffiently studied' (GCVS). The common characteristic of objects in classes I and L is therefore their irregular behaviour. It is therefore understandable that many objects of classes I and L are classified as being members of the other class, not at least since both, class I and L objects are highly heterogeneous and rather loosely assigned to these classes. 

For M type stars we always find a very high classification accuracy (98 \%), regardless of the dataset (original, principal components). This was expected given that these objects have very special lightcurves and very red colours, i.e. they can rather easily been identified. This can also be seen in e.g. Fig.\,\ref{pca11b} where the M type objects form an own `cluster' in the plane of the first two principal components.

There were too few (ordinary) RR-Lyrae objects in the original dataset which is why this class was rejected when building the training/validation template sets. However, there is a large number of RRAB objects (RR stars with asymmetric lightcurves) and we find a high correct classification rate of $\sim$ 93 \% for these objects. 

For class SPB (Slowly Pulsating B stars) almost all objects are classified as ACV variables, the correct classification rate being only $\sim$ 3 \%. This may be a result of the rather similar periodic characteristics and/or colours of these two object types, i.e. the classifier has difficulties to separate these classes.  

Objects of class SRA (semiregular late-type giants) are mainly confused with type L and M stars. Indeed, the correct classification rate is only $\sim$ 6 \% and there is also a considerable spread among the other classes. For SRB stars about half of the objects are classified as I and L stars. A possible reason for this might be the irregular changes in the characteristics of these stars, i.e. they overlap with the class of irregular objects.   

Interestingly, we find from Table\,\ref{origohne} that no object is classified as an ELL (rotating ellipsoidal) variable, not even the `true' ELL stars. Instead, most of these objects are classified as ACV, EB or I type variables.  Though the overall number of objects in the validation set is small (18), the result is possibly significant given that we obtain the same result for the dataset based on the first 14 principal components (see Table\,\ref{pca14ohne}). 
\subsection{How reliable is the training set?}  

In the last Section, we have tried to explain some of the observed (mis)classification  based on the similarity of the objects (or rather their characteristics), i.e. the classification algorithm has difficulties to differentiate between these objects. Especially for classes I and L we found that most of these objects are rather loosely assigned to their object classes in the GCVS. Indeed, it is possible that these classes in the validation (and training!) set are contaminated by many stars of the other class. In this case, the bad classification results are not due to the insufficient classification performance of the classifier (SVM) but rather express the lack of a proper template set for each of these classes.
 
To test this, we validated the SVM on the \textbf{training set} (for the original data set with 51 input features). The results are shown in Table\,\ref{origtrain}. We find an overall classification performance of $\sim$ 78 \% which is significantly better than the $\sim$ 62 \% we obtained for the classification performance based on the validation set (Table\,\ref{origohne}).
However, a close inspection of the numbers reveals that there are many misclassifications, especially among those classes for which we found a bad classification performance for the validation set, too. 

For example, the rate of true positives for class L objects is only $\sim$ 43 \% (see Table\,\ref{origtrain}), even though the classifier was validated on the data set on which it was trained. It is therefore possible that the requirement of mutually exclusive classes is not fulfilled for this data set. Given that some of the objects appear to look very similar (in terms of the features fed to the classifier), a clear distinction is not possible, i.e. the training set is most likely `contaminated' and therefore only of limited value. This, in turn leads to a poor classification model and larger apparent errors. 

\section{Discussion and Conclusion}

In this report we have performed tests on how well we can classify variable objects based on their photometric properties in combination with a supervised classification model. In addition, we tested if a Principal Component Analysis could improve the classification results by reducing the dimensionality of the data set.

Concerning the PCA, we could show that the scaling of the variances of the features (i.e. performing a PCA on the correlation matrix) is appropriate for these data. 

Given that there exists no proper stopping criterium and given that (seemingly) more sophisticated stopping criteria are computationally expensive while not necessarily yielding better results, we decided to do a classification for a set of 14 principal components. The classification results were compared to those based on the original set with 51 features.

We found that there is no significant improvement in using a dataset based on 14 components as compared to the results based on the original dataset. This result is similar to what was found in e.g. \cite{Bailer98} and recently for Gaia photometric data: a PCA does not necessarily improve the classification performance for a highly nonlinear problem. 

Another important result is that the classification accuracy of the specific classifier (here SVMs) cannot be properly assessed given that some of the objects in the template sets (training/validation) cannot be easily assigned to a specific class. This class overlap or `contamination' seems to be a major limitation of the current studies. Future work might want to put more efforts in the construction of a representative template set, although this might prove to be difficult given that certain object types (e.g. L, I) have very similar characteristics. This problem of class definition necessarily leads to a poor model, sets minimum errors and thus degrades the overall performance.

Despite of this limitation, we obtain classification performances of better than 60~\% for nine out of 18 classes. For certain object types (ACV, EA, DCEP, M, RRAB), the fraction of correct classifications (true positives) ranges from 80 to $\sim$ 98~\%. 

Future work will also have to test the classification performance for different signal-to-noise ratios. Concerning PCA, a more sophisticated criteria for selecting the most significant principal components should be used and/or different combinations of several principal components.   
 
Considering Gaia, the level of classification efficiency found in this
document is too low (but keep in mind the above described problems of ill-defined variability classes in the template set). However, it should be noted that in case of Gaia we will be able to define many more
pertinent features of the variability.  Finally, let us remark that this study
is one of the first attempts to study the performances of a classifier
for variable stars in detail.  Once the optimisation of SVMs is well
understood (for a given proper template set), it could be compared to other classification methods. For the Gaia variability analysis, it
is ultimately foreseen that several methods (supervised and unsupervised) are
used simultanously. The choice of these classification methods will
rely on such studies as presented in this work.

\vspace{2cm}
We thank Coryn Bailer-Jones and Torsten Kaempf for helpful discussions.

\vspace{5mm}

\bibliographystyle{apj}
\bibliography{VARI.bib}

\begin{table}
\setlength{\tabcolsep}{1mm}
\begin{sideways}
\parbox[b]{22cm}{
\begin{center}
\begin{tabular}{l||c|c|c|c|c|c|c|c|c|c|c|c|c|c|c|c|c|c}
\cline{1-19}
 & \multicolumn{18}{c}{\textbf{true}} \\
\hline
\hline
\rule[3mm]{0mm}{1mm} \textbf{predicted} & ACV & ACYG & BCEP&  DCEP&  DSCT   & EA &  EB&  ELL&   EW&  GCAS&    I &   L &   M & RRAB  & RS & SPB & SRA & SRB \\             \cline{1-19}                                                            
ACV  &  42  & 2  & 6   &   & 1   &   & 6  & 9  & 1  & 15  & 2  & 1   &    &    &  & 36   &    & \\
ACYG   &   & 4   &    &    &    &   & 2   &    &   & 1  & 2  & 2   &    &    &    &    &    & \\
BCEP  &  2   &   & 3   &    &    &    &    &    &   & 1  & 1   &    &    &    &    &    &    & \\
DCEP   &    &    &   & 65   &   & 1  & 1  & 1   &    &   & 2  & 2   &    &    &    &   & 1   & \\
DSCT  &  1   &   & 2   &   & 35  & 1  & 5  & 1   &   & 1  & 1   &    &   & 2   &   & 1   &    & \\
EA  &  1   &    &    &    &  & 107  & 20   &    &   & 2  & 2   &    &    &    &    &    &    & \\
EB  &  2   &    &   & 1  & 1  & 12  & 60  & 2  & 11  & 11  & 4  & 2   &    &    &    &    &    & \\
ELL   &    &    &    &    &    &    &    &    &    &    &    &    &    &    &    &    &    & \\
EW   &    &    &    &   & 1   &   & 5   &   & 28   &    &    &    &    &    &    &    &    & \\
GCAS  &  5   &   & 1  & 1   &    &   & 2   &    &   & 32  & 3  & 2   &    &    &    &   & 1   & \\
I    &    &    &   & 3  & 1  & 4  & 4  & 3   &   & 5 & 124  & 71  & 1   &   & 1   &   & 1  & 11\\
L    &    &    &   & 2   &    &    &    &    &    &   & 16  & 34   &    &    &    &   & 6  & 16\\
M    &    &    &   & 1   &   & 1   &    &    &    &   & 1  & 1  & 63   &    &    &   & 5   & \\
RRAB   &    &    &    &    &    &    &    &    &    &    &    &    &   & 26   &    &    &    & \\
RS   &   & 1   &    &    &   & 2  & 1  & 1  & 1   &   & 6  & 5   &    &   & 20   &    &    & \\
SPB   &    &   & 1   &    &    &   & 2  & 1   &   & 1   &    &    &    &    &   & 1   &    & \\
SRA   &    &    &    &    &    &    &    &    &    &   & 1  & 2   &    &    &    &   & 1  & 2\\
SRB   &    &    &    &    &    &    &    &    &    &   & 4  & 10   &    &    &    &   & 3  & 23\\
\hline
\hline
\textbf{TP [\%]}  &   \textbf{79.2}  & \textbf{57.1}  & \textbf{23.1}  & \textbf{89.0}  & \textbf{89.7} &  \textbf{83.6} & \textbf{55.6}  &\textbf{0}   &\textbf{68.3}   &\textbf{46.4}  & \textbf{73.4} & \textbf{25.8} & \textbf{98.4}  & \textbf{92.9}  & \textbf{95.2}  &\textbf{2.6}  & \textbf{5.6} & \textbf{44.2} \\ 
\hline
\hline
objects/class  &   53  & 7  & 13  & 73  & 39 &  128 & 108  & 18   &41   &69  & 169 & 132 & 64  & 28  & 21  &38  & 18 & 52  \\
\hline
\end{tabular}
\caption{Confusion matrix for the classification results based on the original dataset and without class weighting. Shown are the numbers of objects in the true versus the predicted classes. For better clarity, only non-zero entries are shown. The line \textbf{TP} shows the number of true positives for each class, i.e. the percentage of correctly identified stars. The last line shows the number of objects per class in the validation set. Summing up the numbers on the diagonal of the confusion matrix and dividing by the total number of objects in the validation set (1071) yields an estimate of the overall classification performance. In this case, we find 62.4~\% of correctly identified objects.  \label{origohne}}
\end{center}
}
\end{sideways}
\end{table}

\begin{table}
\setlength{\tabcolsep}{1mm}
\begin{sideways}
\parbox[b]{22cm}{
\begin{center}
\begin{tabular}{l||c|c|c|c|c|c|c|c|c|c|c|c|c|c|c|c|c|c}
\cline{1-19}
 & \multicolumn{18}{c}{\textbf{true}} \\
\hline
\hline
\rule[3mm]{0mm}{1mm} \textbf{predicted} & ACV & ACYG & BCEP&  DCEP&  DSCT   & EA &  EB&  ELL &  EW&  GCAS &   I  &  L  &  M & RRAB  & RS & SPB  & SRA & SRB \\ 
\hline
ACV  & 45  & 2  & 8   &   & 1   &   & 5  & 9  & 1  & 22  & 3  & 1   &    &    &  & 37   &    & \\
ACYG   &   & 4   &    &   & 1   &   & 1   &    &   & 2  & 2  & 2   &    &    &    &    &    & \\
BCEP  & 1   &   & 3   &    &    &   & 1  & 1   &   & 2   &    &    &    &    &    &    &    & \\
DCEP   &    &    &   & 63   &   & 1  & 1   &    &    &   & 3  & 4   &    &    &    &   & 1   & \\
DSCT  & 1   &   & 2   &   & 33  & 1  & 5  & 2   &   & 1  & 1   &    &   & 2   &   & 1   &    & \\
EA  & 1   &    &    &    &  & 109  & 22   &    &   & 2  & 3   &    &    &    &    &    &    & \\
EB  & 2   &    &   & 2  & 3  & 9  & 62   &   & 16  & 7  & 3  & 2   &   & 1   &    &    &    & \\
ELL    &    &    &    &    &    &    &    &    &    &    &    &    &    &    &    &    &    & \\
EW   &    &    &    &    &    &   & 4   &   & 23   &    &    &    &    &    &    &    &    & \\
GCAS  & 2   &    &    &    &    &   & 1   &    &   & 28  & 4  & 4   &    &    &    &   & 1   & \\
I    &    &    &   & 5   &   & 5  & 2  & 4   &   & 4 & 134  & 84  & 2   &   & 1   &   & 1  & 15\\
L     &    &    &    &    &    &    &    &    &    &   & 7  & 17   &    &    &    &   & 4  & 18\\
M     &    &    &   & 1   &   & 1   &    &    &    &    &   & 2  & 59  & 1   &    &   & 3  & 1\\
RRAB   &    &    &    &    &    &    &    &    &    &    &    &    &   & 24   &    &    &    & \\
RS   &   & 1   &   & 1  & 1  & 2  & 1  & 1  & 1   &   & 5  & 6   &    &   & 20   &    &    & \\
SPB   & 1   &    &    &    &    &   & 3  & 1   &   & 1   &    &    &    &    &    &    &    & \\
SRA    &    &    &    &    &    &    &    &    &    &    &    &    &    &    &    &    &    & \\
SRB    &    &    &   & 1   &    &    &    &    &    &   & 4  & 10  & 3   &    &    &   & 8  & 18\\
\hline
\hline
\textbf{TP [\%]}   & \textbf{84.9}  & \textbf{57.1}  & \textbf{23.1}  & \textbf{86.3} &  \textbf{84.6} & \textbf{85.2}  &\textbf{57.4}   &\textbf{0} &\textbf{56.1}   &\textbf{40.6}  & \textbf{79.3} & \textbf{12.9}  & \textbf{92.2}  & \textbf{92.3}  &\textbf{95.2}  & \textbf{0} & \textbf{0} & \textbf{34.6}\\
\hline
\hline
objects/class  &   53  & 7  & 13  & 73  & 39 &  128 & 108  & 18   &41   &69  & 169 & 132 & 64  & 28  & 21  &38  & 18 & 52  \\
\hline
\end{tabular}
\caption{Confusion matrix for the classification results based on the first 14 principal components. 
The overall classification performance is 59.9~\%. \label{pca14ohne}}
\end{center}
}
\end{sideways}
\end{table}

\begin{table}
\setlength{\tabcolsep}{1mm}
\begin{sideways}
\parbox[b]{22cm}{
\begin{center}
\begin{tabular}{l||c|c|c|c|c|c|c|c|c|c|c|c|c|c|c|c|c|c}
\cline{1-19}
 & \multicolumn{18}{c}{\textbf{true}} \\
\hline
\hline
\rule[3mm]{0mm}{1mm} \textbf{predicted} & ACV & ACYG & BCEP&  DCEP&  DSCT   & EA &  EB&  ELL&   EW&  GCAS&    I &   L &   M & RRAB  & RS & SPB  & SRA & SRB \\    \cline{1-19}                                                            
ACV & 101 &  9 &  10  &  &  1 &  4 &  8 &  8 &  2 &  20 &  2 &  2  &   &   &  &  39  &   & \\
ACYG  & 2 &  24 &  1  &  &  1 &  1 &  4  &   &  &  1 &  7 &  1  &   &  &  1  &   &   & \\
BCEP   &   &  &  22  &   &   &  &  1  &   &  &  1  &   &   &   &   &  &  1  &   & \\
DCEP   &   &   &  &  115  &   &   &   &   &   &   &   &   &   &   &   &   &   & \\
DSCT  & 2  &  &  4  &  &  67 &  1 &  8 &  5 &  1  &   &   &   &   &   &  &  2  &   & \\
EA  &  5  &  &  1  &  &  1 & 318 &  18 &  2 &  1 &  4 &  2  &   &   &  &  2  &   &   & \\
EB  &  2  &   &   &  &  1 &  11 & 155 &  3 &  10 &  7 &  4 &  3  &   &  &  1 &  2  &   & \\
ELL   &   &   &   &   &   &   &  &  7  &   &   &   &   &   &   &   &   &   & \\
EW   &   &   &   &   &   &  &  3  &  &  55  &   &   &   &   &   &   &   &   & \\
GCAS  & 4 &  1 &  5  &   &   &  &  5  &   &  &  92 &  7 &  6  &   &   &   &   &   & \\
I  &  1 &  2  &  &  1 &  1 &  4 &  4 &  1  &  &  2 & 302 & 108 &  1 &  1 &  1  &  &  3 &  20\\
L   &   &   &   &   &   &   &   &   &   &  &  11 &  99  &   &   &   &   &  &  6\\
M   &   &   &   &   &   &   &   &   &   &   &   &  & 125  &   &   &   &   & \\
RRAB   &   &   &   &   &   &   &   &   &   &   &   &   &  &  46  &   &   &   & \\
RS   &  &  5  &   &   &  &  2 &  2 &  2  &   &  &  12 &  5  &   &  &  42  &   &   & \\
SPB   &   &  &  3  &   &   &  &  3 &  1  &  &  2  &   &   &   &   &  &  9  &   & \\
SRA   &   &   &   &   &   &   &   &   &   &   &   &   &   &   &   &  &  21  & \\
SRB   &   &   &   &   &   &   &   &   &   &  &  2 &  3  &   &   &   &   &  &  70\\
\hline
\textbf{TP [\%]}  &  \textbf{86.3}  & \textbf{58.5}  & \textbf{47.8}  & \textbf{99.1} &  \textbf{93.1} & \textbf{93.3}  &\textbf{73.5}   &\textbf{24.1}   &\textbf{79.7}  & \textbf{71.3} & \textbf{86.5} & \textbf{43.6}  & \textbf{99.2}  & \textbf{97.9}  &\textbf{89.4}  & \textbf{17.0} & \textbf{87.5} & \textbf{72.9}\\
\hline
\hline
objects/class    & 117  & 41  & 46  & 116  & 72 &  341 & 211  &29   &69   &129  & 349 & 227 & 126  & 47  & 47  &53  & 24 & 96  \\
\hline
\end{tabular}
\caption{Confusion matrix for the classification results based on the (original) \textbf{training} data. The overall classification performance is  78.0 \% \label{origtrain}}
\end{center}
}
\end{sideways}
\end{table}

\end{document}